\documentclass[12pt]{article}
\usepackage{psfrag,epsfig}
\usepackage{a4,isolatin1}
\usepackage{amsmath,amsfonts,latexsym, amssymb, amscd}

\newtheorem{satz}{Theorem}[section]
\newtheorem{defi}[satz]{Definition}

\newtheorem{bem}[satz]{Remark}
\newtheorem{lemma}[satz]{Lemma}
\newtheorem{koro}[satz]{Corollary}

\newtheorem{conclusion}[satz]{Conclusion}
\newtheorem{ob}[satz]{Observation}

\newtheorem{propo}[satz]{Proposition}

\newcommand{\tit}{\textit}

\newcommand{\N}{\mathbb{N}}
\newcommand{\R}{\mathbb{R}}
\newcommand{\Z}{\mathbb{Z}}
\newcommand{\bewende}{$ \hfill \Box $}

\begin{document}
\thispagestyle{empty}
\begin{center}
\vspace*{1.0cm}

{\large{\bf Wormhole Spaces, Connes' ``Points,\\
 Speaking to Each Other'', and the\\
 Translocal Structure of Quantum Theory}}

\vskip 1.5cm

{\large {\bf Manfred Requardt}}\\email: requardt@theorie.physik.uni-goettingen.de 

\vskip 0.5 cm 

Institut f\"ur Theoretische Physik \\ 
Universit\"at G\"ottingen \\ 
Tammannstr. 1 \\ 
37077 G\"ottingen \quad Germany

\end{center}

\vspace{1 cm}

\begin{abstract}
  We amalgamate three seemingly quite different fields of concepts and
  phenomena and argue that they actually represent closely related
  aspects of a more primordial space-time structure called by us
  wormhole spaces. Connes' framework of non-commutative topological
  spaces and ``points, speaking to each other'', a translocal web of
  (cor)relations, being hidden in the depth-structure of our
  macroscopic space-time and made visible by the application of a new
  geometric renormalisation process, and the apparent but difficult to
  understand translocal features of quantum theory. We argue that the
  conception of our space-time continuum as being basically an
  aggregate of structureless points is almost surely to poor and has
  to be extended and that the conceptual structure of quantum theory,
  in particular its translocal features like e.g.  entanglement and
  complex superposition, are exactly a mesoscopic consequence of this
  microscopic wormhole structure. We emphasize the close connections
  with the ``small world phenomenon'' and rigorously show that the
  micro state of our space-time, viewed as a dynamical system, has to
  be critical in a scale free way as recently observed in other fields
  of network science.  We then briefly indicate the mechanisms by
  which this non-local structure manages to appear in a seemingly
  local disguise on the surface level, thus invoking a certain Machian
  spirit.
\end{abstract} \newpage
\setcounter{page}{1}
\section{Introduction}
We begin our introduction by quoting the following two lucid remarks
by von Weizs\"acker \cite{Weiz}, similar ideas were also entertained
by Wheeler, see e.g. \cite{Wheeler} and some other people. The
quotations are meant to strike the key of our paper.
\begin{quote}
{\small \ldots space-time is not the background but a surface aspect of
reality\ldots It is extremely improbable that this reality
(i.e. quantum reality) will be describable as consisting of events
which are localized in space and time.

The translocal phase relations are ``surplus information'' not lack of
information. Quantum theory knows more, not less, than local classical
physics.}
\end{quote}

The bulk of the following analysis is concerned with an amalgamation
of some important (mathematical) ideas of Connes about a
\tit{non-commutative} generalisation of the \tit{fibration} or
\tit{quotienting-out} principle in e.g. topology and, on the other
hand, of a line of ideas and concepts we developed in recent years in
our approach to quantum gravity. Our analysis results in the
observation of the emergence of a \tit{translocal} component being
hidden in the fine structure of our space-time manifold. We will argue
that this translocal substructure has important consequences for a better
understanding of the many seemingly non-local features of quantum
theory.

We think for example that the famous results of Bell (see \cite{Bell})
are rather an indication that the substructure of quantum theory is
necessarily non-local and not! that the attempts to go beyond the
standard interpretation should be abandoned (as has been also remarked
by Bohm for several times, see \cite{Bohm1},\cite{Bohm2}). We will
advocate the possibility that quantum theory may emerge as an
\tit{effective} theory from a basically translocal microscopic theory
of space-time, its main ingredient being a network of \tit{microscopic
  wormholes}. For that reason we call this class of spaces, we are
going to define and investigate, \tit{wormhole spaces}. That is, we
favor an approach to quantum gravity which considers quantum theory as
a coarse grained consequence of the translocal fine structure of
microscopic space-time, being the consequence of the peculiar
structure of \tit{pregeometry} underlying our ordinary space-time
manifold.

Presently there do exist a variety of routes towards a theory of
quantum gravity, ranging from frameworks, which depart from a more or
less continuous space-time picture, imposing quantum theory more or
less unaltered on the underlying classical structure as an
independent, quasi God-given scheme, to working philosophies which try
to view both quantum theory and gravity as emerging (low-energy)
\tit{effective theories} of a more primordial and basically discrete
substructure (with `discrete' not necessarily meaning `countable' but
rather the absence of a priori continuum concepts). A short and very
incomplete list of papers advocating such more or less discrete
approaches is for example \cite{Fot1} to \cite{Hooft3}.

Our own point of view has been presented in a series of papers in
recent years, to which we refer the reader as to more references and a
more detailed analysis of the different ideas and concepts being
promoted by the various groups sharing this latter working philosophy
(\cite{Req4},\cite{Req1},\cite{Req2},\cite{Quantum}). It goes without
saying that it takes a much longer route to arrive at testable
consequences if one follows this latter more \tit{bottom up oriented}
avenue and before the results can be compared with results derived by
perhaps more immediate \tit{top down} methods (which sometimes are
openly inspired by certain continuum theories such as classical
general relativity). In the following we want to represent such a
testable consequence, that is, predicting the existence of a
microscopic wormhole structure and discussing some of the possible
observable effects. We mention in particular a certain Machian flavor
of this finding.

Furthermore, while some concepts and/or technical tools are shared to
a greater or lesser extent by the various groups working in this
field, this does not necessarily mean that also the respective
frameworks are more or less identical. To give an example, discrete
structures like e.g. (topological) graphs occur of course also
elsewhere in (quantum) gravity research but usually as structures
embedded in a preexisting smooth manifold, sharing sometimes even some
of the metric properties of the ambient space or derive from certain
triangulations, that is, being typically very regular. In contrast to
such scenarios our \tit{dynamic graphs} are strongly fluctuating
irregular dynamical systems (reflecting in a sense the presumed huge
vacuum fluctuations of quantum space-time). Smooth continuum
structures are only supposed to emerge in a sufficently coarse grained
limit which, on the other hand, is only expected to exist provided the
underlying microscopic network is in a very peculiar geometrically
critical state (being closely related to the recently found
\tit{scale-free small world networks}). Therefore it is perhaps
helpful to stress some points which we consider to be characteristic
for our own approach.

We represent the primordial substratum as a densely entangled network
of elementary relations, interactions or information channels, which
is dynamically evolving according to some imposed dynamical law. We
share the wide spread working philosophy that complex and collective
behavior can or should emerge from some simple looking basic laws and
surmise that our space-time or quantum vacuum is no exception. In this
sense our approach is \tit{bottom up}, that is, part of our business
is it to reconstruct the concepts of modern continuum physics as, so
to speak, collective quantities from this primordial substratum (as to
similar ideas cf. for example the last pages in \cite{Wheeler1} or the
philosophy expounded in \cite{Hooft3} or \cite{Hooft2}).  What is
however special in our framework is that \tit{both} the states living
\tit{on} the underlying network and the network architecture itself
are dynamically evolving, the two components interacting with each
other. Hence the interaction of something like ``\tit{geometry}'' and
``\tit{matter}'', which is playing such a dominant conceptual role in
general relativity, is automatically incorporated in nuce in our
approach.  Insofar it transcends the \tit{cellular automaton} approach
advocated by 't Hooft.

Technically we implement this property by introducing \tit{dynamic
  graphs} in which edges can not only reorient themselves under the
dynamics but can also be created and deleted (for the sake of brevity
we refer the interested reader to \cite{Req3} \cite{Small World} or
the above cited earlier papers as to the technical definitions and
some appropriate evolution laws). We note that in \cite{Req3} we
showed in particular that, while having a richer structure, our
networks are also \tit{causal sets}. We avoid however, for the time
being, the discussion of the notorious \tit{problem of time} in
quantum gravity (see for example \cite{Time}). That is, purely for
(technical) convenience we let our network evolve in discrete time
steps, which, however, does not imply that physical time is assumed to
have the nature of an overall discrete clock time.  Our philosophy is
rather that ultimately also physical time will turn out to be a
collective variable emerging on the more coarse grained levels of our
hierarchy of scales of resolution.

We recently came upon the following illuminating remark in
\cite{DeWitt} which beautifully characterizes the continuum version of
the kind of spaces we are going to develop.
\begin{quote}{\small \ldots But if a wormhole can fluctuate out of
    existence when its entrances are far apart \ldots then, by the
    principle of microscopic reversibility, the fluctuation \tit{into}
  existence of a wormhole having widely separated entrances ought to
  occur equally readily. This means that every region of space must,
  through the quantum principle, be potentially ``close'' to every
  other region, something that is certainly not obvious from the
  operator field equations which, like their classical counterparts,
  are strictly local.\ldots It is difficult to imagine any way in
  which widely separated regions of space can be ``potentially close''
to each other unless space-time itself is embedded in a convoluted way
in a higher-dimensional manifold. Additionally, a dynamical agency in
that higher-dimensional manifold must exist which can transmit a sense
of that closeness.}
\end{quote}

While this is not relevant for the following investigation, we hold
the view that, due to our imposed dynamical laws, on a large,
cosmological time scale, our network is in a process of unfolding from
an essentially maximally connected initial state, in which almost all
the degrees of freedom were directly dynamically linked with each
other (in mathematical terms, a simplex or complete graph), towards a
state, which represents our present universe, viz. having a large
classical \tit{diameter} (i.e. lots of possible elementary links being
switched off by the imposed microscopic dynamics), and behaving, at
least macroscopically, in a (quasi-)classical and apparently local
way. The concept of \tit{locality} tells us that sufficiently
separated regions are non- or at most weakly interacting on this
(quasi-) classical level. In a sense, such a classical background is
considered to be the necessary prerequisite for an effective theory
like quantum theory to emerge from a more fundamental theory. On the
other hand we will argue in the following, that there exists, in
addition to the local structure, an almost hidden additional web of
\tit{translocal} (cor)relations between widely separated regions of
classical space-time, which represent, so to speak, the remnants of
this earlier more primordial and much stronger correlated phase (being
prevalent in the big bang era).

These ideas will be expounded and corroborated in sections 5 to 7, the
pivotal section being section 6 in which the structure of the
\tit{critical (scale free) network states} is rigorously analyzed. We
conclude the paper with a brief discussion of the possible
consequences of this peculiar two-level structure of space-time for
the translocal features of quantum theory.

In the following the two papers \cite{Req3} and \cite{Small World} are
of particular technical relevance. In \cite{Req3} we developed in
quite some detail both the conceptual and the numerical machinery for
extracting this mentioned two-level structure from our underlying
network. The framework which makes this possible we called
\tit{geometric renormalisation} or \tit{geometric coarse-graining}.
Its aim is the construction of a non-trivial geometric fixed point
which corresponds to our continuous classical space-time and the
distillation of the necessary preconditions (a \tit{critical,
  scale-free} non-local geometric network state).

These observations immediately lead over to the second paper,
\cite{Small World}. In it we connect our findings with seemingly
closely related observations made in a, at first glance, quite
different context, the so-called \tit{small world phenomenon}
in biological, sociological and other related networks. It turns out
that in both fields we frequently  seem to have roughly two kinds of ties or
links, local ones in closely knitted \tit{friendship neighborhoods}
and non-local ones among only losely connected \tit{acquaintances},
each belonging to a local friendship neighborhood of its own, but
which, typically, do not overlap with each other.In addition to that,
our network  displays an even more peculiar further property which we call a
\tit{wormhole structure}.

We start our investigation by elaborating in sections 2 and 3 on some
of the ideas of Connes, which are later amalgamated with the other
line of our analysis in section 7, the motto being ``\tit{points,
  speaking to each other}''. We note in particular, that in our
approach \tit{physical points} (also called \tit{lumps} by us, cf.
\cite{Req2}) have a rich internal structure. It is therefore
interesting that the limits of the classical point-concept are also
clearly felt in pure mathematics, see for example the beautiful essay
by Cartier, \cite{Cartier}, who remarks:
\begin{quote}
{\small \ldots The central problem is that of the points of
  space\ldots\\
\ldots the only things that matter are their mutual
relationships\ldots\\
\ldots To a given order the infinitesimals of the immediately higher
order appear to be points without structure, until we open the box
that they constitute and that reveals infinitesimals of a higher order
playing provisionally the role of points.}
\end{quote}

While our model systems, when appropriately \tit{coarse-grained}, can
presumably lead to something like classical gravity in a low-energy
limit, we say however almost nothing about this important point in
the present paper. We only remark that it is obvious that concepts
like curvature, dimension and the like are contained in our approach
(as to dimensional concepts see e.g.  \cite{dim}, \cite{Req3} or
\cite{Small World}).

To give an example, a concept closely related to the notion of
curvature, i.e. the relation between the number of points lying in a
surface, having a certain fixed distance from a given point and the
distance itself, can easily be formulated in our network approach. On
the one hand remember the famous thought experiment of Einstein of a
rotating reference frame (leading to the conclusion that space-time is
non-euclidean, cf. \cite{Einstein1},\cite{Einstein2} or
\cite{Moeller} sect.8.3). On the other hand, we studied such a relation in
\cite{dim} or \cite{Small World} to introduce the notion of (fractal)
graph dimensions. In network or graph theory it is called the
\tit{distance degree sequence relative to a vertex}. Concepts like the
above can also be discussed in the more abstract and wider setting of
\tit{metric spaces} (see the interesting ideas of Gromov in
\cite{Gromov} or \cite{Bridson}).Our graphs and networks are natural
examples of such (discrete) metric spaces. As our graphs and their
coarse grained descendants (lump spaces) are even \tit{geodesic metric
spaces}, the same will hold for the continuous limit manifold if it
exists. It is then an extremely interesting question under what
conditions this limit manifold carries a Riemannian metric (a problem
already envisaged by Riemann him self! \cite{Riemann}).
\section{\label{Points}Physico-Mathematical Aspects of Point-Set\\ Topology}
One of the ideas of Connes is to give the \tit{interior of points},
which, on their side, frequently result from some sort of contraction
or identification of subensembles of finer constituents, a non-trivial
noncommutative structure.

A large part of modern physics still relies on the ideas of
mathematical continuum geometry and point-set topology. Furthermore,
in most fields of physics, with the exception of general relativity,
space typically occurs as some fixed background structure, not
participating in the dynamics of the constituents of matter. But even
in general relativity space-time is contrived as a preexisting
topological manifold of structureless points which rather play the
role of labels of events. In some sense this is a slightly dubious
point of view as no one has ever seen these individual points and
without coordinate systems and events it would be hard to tell the
individual space-time points from each other anyhow.

This manifold is a dynamic agent in general relativity but not so much
as a dynamical system (as in our approach) with direct interaction
between the constituents, viz, the points. This interaction is rather
mediated by matter-fields and/or the metric tensor or connection
fields. These are considered to be quantities being attached to the
points, but the points themselves appear to be unaffected by the
dynamics, nor are these fields usually regarded as actually encoding
the internal structure of points or their infinitesimal neighborhoods.
In this sense the points of the manifold are \tit{ideal elements} in a
twofold way (as to a discussion of the notion of ideal concepts see
e.g. \cite{Quantum} and further references there).  They neither do
act nor are acted upon, they serve only as carriers of fields.
Relating this field approach to our point of view; as our (physical)
points are little densely knitted subunits, they are capable of
carrying internal states. From a more macroscopic point of view one
may regard these states as being located in the \tit{infinitesimal}
neighborhood of the points of the space-time continuum. It is then
suggestive to view fields \tit{at} points as encoding in a
coarse-grained way the fine structure of the unresolved microscopic
lumps. The same applies to theories of the Kaluza-Klein type.

While on the physical side, this clean picture is a little bit blurred
by the advent of quantum theory, with classical concepts of localized
objects and points now becoming slightly obscure, nevertheless, the
whole framework is still, cum grano salis, moulded in this universal
conceptual form of continuous spaces and local fields living on them.

A concept like interaction between points played also no notable role
in classical mathematics (apart, perhaps, from graphs, to which we
come below). However, there exists the widespread concept of
\tit{identification} or \tit{quotienting out}, that is, with $\pi$ a
surjective map
\begin{equation}\pi:X\to Y\end{equation}
$X$ and $Y$ two spaces, we can decide to identify the set of points,
lying in the preimage of $y\in Y$ with $y$ and, by the same token, the
partitioning of $X$ by $\pi$ with $Y$. We write
\begin{equation}X/\!\!\sim=Y \end{equation}
the equivalence relation being induced by $\pi$. Correspondingly we
can introduce the \tit{quotient space} of $X$ by $\sim$ if we are
given an equivalence relation on $X$. Physically we can equally well
view the map as a \tit{sorting} of the points of $X$ by the points of
$Y$ or as imposing a value property on $X$.

Ordinarily, the individual points in the respective equivalence
classes are then identified, that is, in general the emerging
structure becomes poorer or coarser. As stressed by Connes (see the
following two sections) the structure can in fact become so poor and
coarse in many relevant cases as to become virtually void and
uninteresting (while, on the other hand, the underlying fine structure
may be extremely complicated and far from trivial). It was an
important observation of Connes that in such extreme situations the
fine structure of such \tit{leaf-} and \tit{identification spaces} can
be more appropriately encoded in a \tit{noncommutative} structure,
living \tit{over} such spaces (or rather, certain extensions being
associated with such $X/\!\sim$).

This is the mathematical aspect. As to physics, Connes in
\cite{Connes2} made the subtle remark as to such identified points,
$\{a,b\}$ of some initial space $X$: ``\ldots to allow them to `speak'
to each other ''. In physical terms, one may interpret this as
\tit{interaction} among the points of a space or manifold.

Orbits, leaves and other subset structure, occurring in the construction of
quotient spaces, are examples of equivalence relations. It turns out
that, in our approach, this is a too narrow framework. An equivalence
relation is a subset, $R\subset X\times X$, $X$ a certain set, with
the following properties called, reflexivity, symmetry and
transitivity, respectively, i.e.
\begin{equation} \forall x\in X:\,(x,x)\in R\,,\,(x,y)\in R\rightarrow
  (y,x)\in R\,,\,(x,y),(y,z)\in R\rightarrow (x,z)\in R\end{equation}
Other types of relations are, for example, \tit{adjacency}: $R$ is symmetric
and irreflexive ($(x,x)\not\in R$),\tit{partial
order}: $R$ is reflexive, antisymmetric ($(x,y)\in R\rightarrow
(y,x)\not\in R$) and transitive and so forth.

In our network approach the type of relations which naturally occur
are even more special. They are generalisations of adjacencies. Our
dynamical systems are assumed to live on (simple) graphs. If these
graphs are unoriented they define an adjacency, with edges between the
vertices denoting the (symmetric) relation. If we take, in addition
the dynamical laws into account, our graphs become oriented graphs,
with the orientation being (clock)time dependent. That is, at each
instant of time the edges or the relations happen to be oriented anew
so that either $(x,y)$ or $(y,x)$ occurs in $R(t)$. Put differently,
the type of relations which are also relevant in the following are
irreflexive and antisymmetric. At each instant of time and for each
vertex $x$ we have a subset $[x]_{in}$ of $X$ influencing $x$ and a
subset $[x]_{out}$ being influenced by $x$, the first set given by
edges having $x$ as target, the latter set being given by edges having
$x$ as source.  It is important that, typically, these relations are no
longer transitive.

For convenience we always assume that in the relation $R$ every $x\in
X$ occurs as a possible first entry in $(x,y)$, put differently, each
$x$ is related to at least one other element of $X$. In the directed
case this means that each $x$ has at least one \tit{outgoing edge}.

If the base set $X$ is countable, we can represent these relations by
(in general) non-symmetric \tit{adjacency matrices}. Labelling the
rows and columns by the members of $X$, the corresponding entries in
the row belonging to $x$ are either $1$ if the respective column label
belongs to $[x]_{out}$ or $0$ else. Correspondingly, the column
labelled by $x$ has entries with value $1$ at the places belonging to
$[x]_{in}$. In this way the wiring structure of the oriented graph can
be neatly encoded in a matrix. A more detailed analysis of properties
of such matrices and (directed) graphs was made in \cite{Req5}.

\section{A Road to Noncommutative Spaces}
As perhaps not everybody is familiar with the details of the
mathematical constructions we briefly review this topic in the
following in our own words.
\subsection{Mathematical Prerequisites and Motivation}
An important conceptual tool in modern mathematics to construct new
spaces from given ones, is the quotient operation, that is, dividing a
bigger point set by an \tit{equivalence relation}. Starting from a
set, $X$, and a particular subset, $R\subset X\times X$, having the
above properties of an equivalence relation we form a new space
denoted by $X/R$, $X/\!\!\sim$ or simply $\tilde{X}$, with points
being the equivalence classes, $\tilde{x}$, defined by $R$, i.e.
\begin{equation}y\in \tilde{x}\;\text{if}\;(x,y)\in R\;\text{and}\; \tilde{y}=\tilde{x}\;\text{if}\;y\in \tilde{x}    \end{equation}

If $X$ carries a \tit{topology} we can endow the new space with the
canonical \tit{quotient topology}, being the finest topology on
$\tilde{X}$ so that the quotient map
\begin{equation}\pi:X\to\tilde{X}\quad,\quad x\to \tilde{x}       \end{equation}
is continuous. In other words, a set, $\tilde{O}\subset\tilde{X}$ is
open iff $\pi^{-1}(\tilde{O})$ is open in $X$.

Typical cases in point are identification or quotient spaces derived
from the action of a group, $G$, on $X$, the equivalence classes being
the orbits of the group action, i.e.
\begin{equation}\tilde{x}:=\{g\cdot x,g\in G\}\end{equation}
In this case each $g$ is assumed to act as a permutation or bijection
on $X$, that is, the space $X$ is partitioned (or \tit{foliated}; at
the moment we do not intend to give the precise definition, see
e.g. \cite{Connes1} or \cite{Tamura}) into \tit{orbits} or \tit{leaves}.

In most of classical mathematics, the quotient spaces being studied
typically carry a non-singular (e.g. Hausdorff-) quotient topology. On
the other hand, as strongly emphasized by Connes, there do exist lots
of interesting (quotient) spaces with highly irregular or fragmented
orbits, leaves or partitionings. A consequence may be that the
ordinary induced topology is trivial, the only open or closed sets
being the total space and the empty set, called the \tit{indiscrete}
or \tit{coarse} topology. It follows that the associated function
spaces are also trivial, consisting only of constant functions (in
case the space is connected).

In other words, the ordinary \tit{commutative} philosophy, encoding
the topology of quotient spaces in the corresponding function algebra
over the space (via the Gelfand-isomorphism), turns out to be
completely insufficient as the space may, nevertheless, have an
extremely rich internal structure, which is, to express it in physical
terms, no longer resolved by the microscope, given by the associated
function algebra.

Various paradigmatic examples are discussed in the book of Connes
(\cite{Connes1}). A nice review is also \cite{Connes2}. The presumably
most thoroughly studied example is the so-called \tit{noncommutative
  torus} (NCT), (\cite{Connes1},\cite{Connes2} or, as to the purely
mathematical aspects, \cite{Rieffel1} or \cite{Varilly}). A
pedagogical review, more adressed to theoretical physicists, is for
example \cite{Bigatti1}.

The model itself has already been known in classical mechanics for a
long time in connection with ergodic theory (Kronecker foliation, see
\cite{Arnold} p.72ff). With coordinates on the two-torus, $T^2$, given
by
\begin{equation}(2\pi\cdot x,2\pi\cdot y)\quad,\quad 0\leq
  x,y<1\end{equation}
or, equivalently
\begin{equation}T^2\cong \R^2/\Z^2   \end{equation}
as topological quotient space, one studies the rotation map
\begin{equation}\dot{x}(t)=\alpha_1\quad ,\quad \dot{y}(t)=\alpha_2 \end{equation}
If $\alpha_1/\alpha_2$ is rational, the induced leaf space, that is
$T^2/\!\!\sim$, is a nice topological space in the sense discussed
above, as the orbits
\begin{equation}(x(t),y(t))\;,\; x(0)=x_0,y(0)=y_0   \end{equation}
close on themselves after a finite number of cycles.

The situation changes drastically for $\alpha_1/\alpha_2$ irrational.
In that case all the orbits of the flow are dense in $T^2$. This is a
consequence of the \tit{Poincar\'e recurrrence theorem} (see
\cite{Arnold}). This results in a degeneration of the canonically
defined quotient topology on $T^2/\!\!\sim$ to the indiscrete topology
(for a more detailed discussion of topological questions see
\cite{Cantor-Connes}).

As a consequence the algebra of continuous functions on $\tilde{X}$
degenerates to the constant functions (for $\tilde{X}$ being
connected). A parallel result holds for measure theory, based on
Borel-measures. The reason is the following. For $\alpha_1/\alpha_2$
irrational, the corresponding flow is ergodic. Employing the canonical
quotient measure, induced by Lebesgue measure on $\R^2$, measurable
functions on $\tilde{X}$ are functions, being invariant on the leaves.
That is, the pull back leads to functions, being invariant under the
flow. However we have the important result that invariant measurable
functions under an ergodic flow are constant on any set of full
measure. Analogously, any invariant measurable set is either of zero
measure or meassure one (for normalized measure spaces), see
\cite{Sinai}.

In our particular case this can be visualized as follows. The
individual leaves are of course measurable but have Lebesgue measure
zero. If we want to have a set with non-vanishing measure, we may for
example choose a full interval of initial conditions for the flow. As
the flow is ergodic, the corresponding invariant set is however the
full 2-torus.
\\[0.3cm]
Remark: We note in passing (without discussing this possibly
interesting point in more detail at the moment), that one may
introduce measures of the \tit{fractal type} on $\tilde{X}$, leading
to a larger class of measurable functions. The relevance for
(continuum) physics is however not immediately obvious.\vspace{0.3cm}

From the above discussion it follows that the functorial
identification of (topological) spaces and abelian algebras becomes
obsolete in these (not so infrequent) situations. In the past such
spaces have mostly been studied in a more algebraic manner. In the
following we want to adopt a slightly more geometric (or topological)
point of view and emphasize aspects which will, hopefully, exhibit the
relation to our own approach to \tit{quantum space-time physics}.

To do this, a closer inspection of the arguments and ideas, given by
Connes in for example the first chapter of \cite{Connes1}, called
``Noncommutative spaces and Measure Theory'' and in particular the
subsection I.4: ``Geometric Examples of von-Neumann algebras'', is
helpful.
\subsection{Noncommutative Quotient Spaces, the Construction}
It is important for the following to understand in more detail some of
the technical subtleties, underlying the construction of operator
algebras on e.g. leaf spaces, given in \cite{Connes1}. We will see
that, strictly speaking, the ``noncommutative'' construction is
actually performed over a particular \tit{fiber-bundle} with \tit{base
  space} $X$ and \tit{not} really over the singular quotient space
$\tilde{X}$, which rather plays an intermediary role by supplying the
fibers over the points of $X$.

We simplify the discussion by assuming the leafs, or more generally,
equivalence classes of points, to be countable sets. Cases in point
are e.g. the action of a discrete group, $G$, on a manifold, $V$. We
assume the underlying space, $X$, to be a measure space. In the more
general (non-countable) case one may take the \tit{Lebesgue measure
  class} and deal (in the absence of a canonical volume form) with the
Hilbert space of \tit{half-densities} or \tit{half-forms} (cf.
\cite{Connes1} or \cite{Abraham}).

With $\tilde{X}$ having no longer an interesting structure as a
measure space, we, following Connes, proceed in the following way. We
take the initial space, $X$, a manifold say, and errect a \tit{Hilbert
  bundle}, $\tilde{H}$, over $X$ by attaching, in an intermediate
step, to each point, $x\in X$, the corresponding equivalence class,
$\tilde{x}$ (orbit, leaf) thus forming the subsets $(x,\tilde{x})$ in
$X\times X$. Note that this implies that now all the points,
$x_i,x_j$, belonging to the same leaf, carry the \tit{same} fiber,
$\tilde{x_i}=\tilde{x_j}$.

With the fibers being countable, we then errect over each fiber,
$\tilde{x}$, the $l_2$-Hilbert space, $H(x)$, of sequences
\begin{equation}\{\tilde{f}(x_i)\}\;,\;x_i\in\tilde{x}\;,\;\sum_{\tilde{x}}|\tilde{f}(x_i)|^2<\infty\end{equation}
with a basis consisting of the functions
\begin{equation}\tilde{f}_j\;,\;\tilde{f}_j(x_i)=\delta_{ij}   \end{equation}
That is, instead of the singular space, $\tilde{X}$, we study the
Hilbert bundle, $\tilde{H}$, with base space the nicer space, $X$, and
fibers being the $l_2$-spaces over the leaves, $\tilde{x}$, indexed by
$x\in X$.\\[0.3cm]
Remark: Note that all the Hilbert spaces are isomorphic (but not!
canonically isomorphic) to a standard $l_2$-space, which may be
regarded as \tit{standard fiber}.\vspace{0.3cm}

It is now easy to construct measurable sections over $X$ in the
following way. Pick a $\tilde{f}_x$ in each $H(x)=H(\tilde{x})$, the
index, $x$, running in $X$, \tit{not} in the fiber over $x$. By the
same token, this defines a function, $f$, over the space $R\subset
X\times X$, $R$ given by the equivalence relation or foliation,
$(x,y)\in R$ if $\tilde{x}=\tilde{y}$.
\begin{defi}The section, $\tilde{f}$, of Hilbert vectors, $\tilde{f}(x)=\tilde{f}_x$, is called
  measurable if the corresponding induced function, $f$, is measurable
  over $R\subset X\times X$ with $f$ given by
\begin{equation}f(x,y):=\tilde{f}_x(y) \end{equation}
i.e. the Hilbert vector $\tilde{f}_x$ evaluated at element $y$ in
$\tilde{x}$. In the same sense we define square integrable sections
over $X$ with values in $H(x)=H(\tilde{x})$. We denote this space by
$L^2(\tilde{H})$.
\end{defi}

It is important to note that this Hilbert bundle is a bundle over $X$
and that even if $x,y$ belong to the same fiber, i.e.
$\tilde{x}=\tilde{y}$, the Hilbert vectors in $H(x)\,,\,H(y)$ can be
independently chosen, that is, $\tilde{f}_x\neq \tilde{f}_y$ in
general. Consequently, this structure alone does \tit{not} yet reflect
the true leaf structure of $\tilde{X}$. On the other hand, the above
amplification construction is technically necessary due to the, in
general, degenerated structure of $\tilde{X}$. The leaf structure will
be encoded in the operator algebra constructed below.

It was realized by Connes that we can both get an interesting
mathematical structure and a characterisation of the underlying
singular leaf space by now taking the natural operator algebras or
matrix algebras of bounded operators on $H(x)$ \tit{leafwise}, that
is, we define an operator valued function, $\tilde{A}$, over $X$ with
the proviso
\begin{equation}X\ni
  x\to\tilde{A}_x=A_{\tilde{x}}\quad\text{i.e.}\quad
  \tilde{A}_x=\tilde{A}_y\quad\text{if}\quad \tilde{x}=\tilde{y} \end{equation}
with $A_{\tilde{x}}$ a bounded operator in $H(\tilde{x})$.\\[0.3cm]
Remark: Connes calls such operators \tit{random operators}.\\[0.3cm]
\begin{defi}We call such a section of operators,$\tilde{A}$,
  measurable, if for any pair of measurable Hilbert vector
  sections, $\tilde{f}\,,\,\tilde{g}$, $(\tilde{f}_x|\tilde{A}_x\cdot
  \tilde{g}_x)$ is measurable.
\end{defi}
\begin{lemma}The random operators with norm given by
  $\operatorname{ess\,sup}\|(\tilde{A}_x)\|$ bounded, form a
  von-Neumann algebra over $L_2(\tilde{H})$ under pointwise multiplication.
\end{lemma}
(see \cite{Connes1}). We now see that these random operators or the
corresponding von-Neumann algebra characterizes the leaf structure in
a particular \tit{noncommutative} way.
\section{The Network of Interacting Points}
\subsection{The Underlying Network $QX$}
In section \ref{Points} we argued in favor of a framework which
implements the interaction among the points of a manifold and that
this is, on the other hand, inherent in some of the ideas of Connes.

The model system we start from is a dynamic discrete graph or network
assumed to emulate crucial aspects of (quantum) space-time on the
Planck scale. With the help of a coarse-graining or \tit{geometric
  renormalisation} process (\cite{Req3}) we undertake to construct a
macroscopic \tit{fixed point} representing our continuous space-time
on the macroscopic or mesoscopic level. But in contrast to the
ordinary continuum employed in classical mathematics or physics, which
is assumed to behave in a purely local way, in our approach, the
continuum limit develops quasi automatically (brought to light by the
renormalisation construction) an intricate and largely hidden extra
(non-local) structure among its points.

For notational convenience we only introduce some notation in the
following and refer the interested reader to \cite{Req3} or
\cite{Small World} for more technical details.
\begin{defi}\label{graph1}A simple, countable, labelled, (un)directed
  graph, $G$, consists of a (countable) set of nodes or vertices, $V$,
  and a set of edges or bonds, $E$, each edge connecting two of the
  nodes.  For convenience there exist no multiple edges (i.e. edges,
  connecting the same pair of nodes) or elementary loops (a bond,
  starting and ending at the same node). In this situation the bonds
  can be described by giving the corresponding set of (un)ordered pairs
  of nodes. The members of $V$ are denoted by $x_i$, the bonds by
  $e_{ij}$, connecting the nodes $x_i$ and $x_j$. 
\end{defi}
Remark: The assumption of a countable vertex set is only made for
technical convenience. We could also admit a non-countable vertex set.
From a physical point of view one may argue that the \tit{continuum}
or uncountable sets are idealisations, anyhow.\vspace{0.3cm}

We note that graphs carry a natural structure which can be employed to
emulate the interaction between the nodes or points.This becomes more
apparent if we impose dynamical network laws on these graphs such that
they become discrete dynamical systems.  Henceforth we denote such a
dynamical network, which is supposed to underly our continuous
space-time manifold, by $QX$ (``\tit{quantum space}'').  We omit a
more detailed discussion of the dynamical evolution of states on
graphs or networks, which can be found in our above mentioned papers
as we want to concentrate primarily on the emergent non-local aspects
of our model systems. We only want to emphasize the following point.

It is important that in our approach the bond states are also
dynamical degrees of freedom which, a fortiori, can be switched off or
on.  Therefore the \tit{wiring}, that is, the pure \tit{geometry} (of
relations) of the network changes constantly and is hence also an
emergent, dynamical property, \tit{not} given in advance in form of
some static background geometry. Furthermore, in the network laws we
have studied so far, the individual edges carry states which can take
the values $\pm 1,0$ and which are naturally associated, via the
dynamical law, with the two possible orientations of the edge or its
silent, inactive state respectively (in the graph framework the edge
is simply considered to be temporarily absent if the edge state is
zero). These local states are updated after every clock-time step
(depending on the state of the nodes in the local environment).
\\[0.3cm]
Remark: One sees from this that our \tit{cellular networks are
  generalisations} of the more common but also geometrically more
rigid \tit{cellular automata}.\vspace{0.3cm}

Consequently, the nodes and bonds are typically \tit{not} arranged in
a more or less regular array, a lattice say, with a fixed
near-/far-order. It is remarkable that similar ideas have also been
entertained in the theory of \tit{cellular automata}, where systems
have been studied which selforganize, in a dynamical process, their
lattice structure (see the beautiful book of Ilachynski, \cite{Ila}).
\subsection{\label{cliques} Dynamical Networks as Random
  Graphs} As we are dealing with very large graphs, which are, a
fortiori, constantly changing their shape, that is, their distribution
of active bonds, we make the assumption that the dynamics is sufficiently
stochastic so that a point of view may be appropriate, which reminds
of the working philosophy of \tit{statistical mechanics}.

It was recently argued (\cite{Small World} and further literature
cited there) that the random graph framework may be too narrow to
fully reproduce the observed \tit{near-, far-order} of so-called
\tit{scale-free small world networks} which seem to be the crucial
prerequisites for the emergence of a non-trivial critical continuum
fixed point of our coarse graining process. But nevertheless the
random graph picture is still the natural starting point and the
basis of a perhaps more advanced theory.

Visualizing the characteristics and patterns being prevalent in large
and ``typical'' graphs was already a notorious problem in
\tit{combinatorial graph theory} and led to the invention of the
\tit{random graph} framework (\cite{Req3}, \cite{Small World}). The
guiding idea is to deal with graphs of a certain type in a
probabilistic sense, that is, forming a probability space with
elementary events certain graphs. This turns out to be particularly
fruitful as many graph characteristics (or their absence) tend to
occur with almost certainty in a probabilistic sense (as has been
first observed by Erd\"os and R\'enyi). In the following we are
dealing with random graphs living over a fixed node set, having the
independent \tit{edge probability} $0<p \leq 1$. The probability that
a particular graph, $G_m$, with $m$ edges occurs is thus
\begin{equation}pr(G_m)=p^m(1-p)^{N-m}\end{equation}
with $n,N=n(n-1)/2$ the number of nodes, the maximal possible number
of edges respectively. The standard source is \cite{Bollo1}.

In the above papers (in particular in \cite{Req3}) we mainly
concentrated on properties of so-called \tit{cliques}, their
statistical distribution (with respect to their order, $r$, i.e.
number of vertices), degree of mutual overlap etc. We then studied
these properties in the consecutive stages and phases of our
renormalisation process, being associated to the various levels of
magnification or resolution of our space-time manifold.
\begin{defi}[Subsimplices and Cliques]With $G$ a given fixed graph and
  $V_i$ a subset of its vertex set $V$, the corresponding {\em induced
    subgraph} over $V_i$ is called a subsimplex or {\em complete
    subgraph}, if all its internal pairs of nodes are connected by a
  bond. In this partially ordered set there exist certain {\em maximal
    subsimplices}, that is, every addition of another node (together
  with the respective existing bonds to other nodes of the subset)
  destroys this property.  These maximal simplices are called {\em
    cliques} in combinatorics and are the candidates for our {\em
    physical proto-points}.  Henceforth we denote them by $S_i$.
  \end{defi}
  
  For the underlying reason why we concentrated on this particular
  graph characteristic we provided some motivation in the above cited
  papers. To put it briefly, we will associate these cliques with the
  nested structure of lumps (or physical points) making up our
  ordinary space-time.  That is, the cliques are assumed to look like
  ordinary points under low magnification but show their internal
  (infinitesimal) nested structure under sufficiently high resolution
  (cf. the remarks by Cartier cited in the introduction). In
  \cite{Small World} we related these ideas to astonishingly similar
  ideas in a, at first glance, quite unrelated field where the local
  lumps are called \tit{friendship neighborhoods} or \tit{clumps} and
  the non-local ties \tit{aquaintances}.
  
We can introduce various \tit{random function} on the above
probability space.  For each subset $V_i\subset V$ of order $r$ we
define the following random variable:
\begin{equation}X_i(G):=
\begin{cases}1 & \text{if $G_i$ is an $r$-simplex},\\  
 0 & \text{else}
\end{cases}
\end{equation}
where $G_i$ is the corresponding induced subgraph over $V_i$ in $G\in
{\cal G}$ (the probability space). Another random variable is then the
\tit{number of $r$-simplices occurring in a given $G$}, denoted by
$Y_r(G)$ and we have:
\begin{equation}Y_r=\sum_{i=1}^{\binom{n}{r}}X_i\end{equation}
with $\binom{n}{r}$ the number of $r$-subsets $V_i\subset V$. With respect
to the probability measure introduced above we have for the
\tit{expectation values}:
\begin{equation}\langle Y_r \rangle = \sum_i \langle X_i \rangle\end{equation}
and
\begin{equation}\langle X_i \rangle = \sum_{G\in{\cal G}} X_i(G)\cdot
  pr(G_i=\text{$r$-simplex in}\;G).\end{equation} 

For $\langle Z_r \rangle$, the {\em expected number of $r$-cliques}
(i.e. maximal! $r$-simplices) in the random graph, we have then the
following relation
\begin{equation}\langle Z_r
  \rangle=\binom{n}{r}\cdot(1-p^r)^{n-r}\cdot
  p^{\binom{r}{2}}\end{equation}

This quantity, as a function of $r$ (the \tit{order} of the
subsimplices) has quite a peculiar numerical behavior. We are
interested in the typical \tit{order of cliques} occurring in a
generic random graph (where typical is understood in a probabilistic
sense.
\begin{defi}[Clique Number]The maximal order of occurring cliques contained in
  $G$ is called its {\em clique number}, $cl(G)$. It is another random
  variable on the probability space ${\cal G}(n,p)$.
\end{defi}
It is remarkable that this value is very sharply defined in a
typical random graph. Using the above formula for $ \langle
Z_r\rangle$, we can give an approximative value, $r_0$, for its
expectation value and get
\begin{equation}r_0\approx 2\log(n)/\log(p^{-1})+ O(\log\log(n))\end{equation} (cf. chapt. XI.1 of \cite{Bollo1}). It holds that practically all the
occurring cliques fall in the interval $(r_0/2,r_0)$ (for a
quantitative and numerical discussion see \cite{Req3}).  We believe
that the random graph picture will reproduce at least the qualitative
behavior of such extremely complex dynamical systems, being well aware
of the possible limitations and necessary generalisation of this
picture ( \cite{Small World}).
\section{The Geometric Coarse-Graining or Renormalisation Process}
We now are going to set up the connection between the two fields
discussed in the preceding sections. That is, on the one hand, the
concept of nasty (quotient) spaces having very erratic and in some
cases dense orbits, leaves etc., and, on the other hand, our hierarchy
of cellular networks or dynamic graphs, which emerges from a certain
kind of geometric renormalisation group. As the construction of this
nested structure of lumps within lumps was discussed in quite some
detail in \cite{Req3} and the emergence of the small world effect in
\cite{Small World} we can be relatively brief.

The idea to construct a macroscopic (quasi) continuum from an
underlying more erratic and discrete primordial substratum via some
sort of coarse graining is, as we think, not unnatural (see in
particular \cite{Req2} and our discussion of earlier work by Menger et
al about \tit{random metric spaces}). These considerations led to our
concept of a geometric renormalisation group. We start with a graph,
$G$, and, in a first step, pick up its cliques promotingt them to the
(meta) nodes of a coarser graph, the so-called \tit{clique graph}, $G_{cl}$. If
these cliques are not too small, it should make a physical difference
whether two selected cliques have an appreciable overlap of common
nodes, if this overlap is only marginal (very few common nodes) or
even empty.

In \cite{Req3} realistic numerical examples were studied in which the
typical clique size was roughly of order $10^3$. The random graph
framework allowed to calculate the probability distribution of
expected clique overlaps and related graph characteristics. The
quantitative calculations are however relatively tricky and involved.
In the case of a strong overlap the interaction between the respective
cliques is more intensive while in the latter cases it is weak and/or
indirect, that is, the internal state of the other clique is only
feebly felt if the interaction is weak. It is the merit of the
renormalisation group that it clearly separates these two different kinds of
interaction after several coarse graining or renormalisation
steps.

Our above described procedure suggests the physical assumption that
\tit{classical} macroscopic behavior is hoped to emerge if we neglect
the fine details on small scales (e.g. fluctuations). That is, we only
will draw a link between two cliques or lumps, $S_i,S_j$, if the
common overlap is non-marginal compared to the typical order of the
cliques on the respective renormalisation level. The graph, thus
constructed, we call the \tit{purified clique graph} relative to $G$.
To put it more succinctly:
\begin{itemize}
\item Starting from a given fixed graph, $G$, pick the (generic) \tit{cliques},
  $S_i$, in $G$, i.e. the subgraphs, forming maximal subsimplices or
  cliques in $G$ with their order lying in the above mentioned
  interval, $(r_0/2,r_0)$.
\item These cliques form the new nodes of the \tit{clique-graph},
  $G_{cl}$ of $G$.New bonds in the clique graph are drawn between
  cliques provided they have a sufficient overlap.
\end{itemize}
\begin{bem}The random graph framework shows the highly non-trivial
  fact that practically all occurring cliques have a number of nodes lying
  in the above interval, that is, have a typical size.
\end{bem} 

We repeat this process of going from a graph to its purified clique
graph sufficiently many times until we arrive at a (quasi-)continuous
manifold, emerging as a fixed point of our renormalisation process
(and being reflected by the emergence of a quasi-static regime in
which the graph structure do no longer change appreciably in the
consecutive coarse-graining steps). Such a macroscopic fixed point can
however only expected to emerge provided the original network has been
in a very peculiar, i.e., (quasi-)\tit{critical} state as has been
described in section 8 of \cite{Req3}.

On each level of coarse-graining, that is, after each renormalisation
step, labelled by $l\in\Z$, we get, as in the block spin approach to
critical phenomena, a new level set of cliques or lumps,$ S^l_i$, ($i$
labelling the cliques on renormalisation level $l$), consisting on
their sides of $(l-1)$-cliques which are the $l$-nodes of level $l$,
starting from the level $l=0$ with $G=:G_0$. That is, we have
\begin{equation}S^l_j=\bigcup_{i\in j}
  S_i^{(l-1)}\;,\;S_i^{(l-1)}=\bigcup_{k\in i}
  S_k^{(l-2)}\;\text{etc.}\end{equation}
($i\in j$ denoting the $(l-1)$-cliques, belonging, as meta nodes,
 to the $l$-clique, $S_j$).
These cliques form the meta nodes in the next step.
\begin{defi}The cliques, $S_i^0$, of $G=:G_0$ are called zero-cliques. They
  become the one-nodes, $x_i^1$, of level one, i.e. of $G_1$. The one-cliques,
  $S_i^1$, are the cliques in $G_1$. They become the 2-nodes, $x_i^2$,
  of $G_2$ etc.  Correspondingly, we label the other structural
  elements, for example, 1-edges, 2-edges or the distance functions,
  $d_l(x_i^l,x_j^l)$. These higher-level nodes and edges are also
  called meta-nodes, -edges, respectively.
\end{defi}

If we collapse these new cliques to meta-nodes we do no longer see
their internal structure. If, on the other hand, we decide to keep
track of their internal organisation (cf. also the remarks of Cartier
in the introduction) we have the following (Russian doll like) picture
(where for the sake of graphical clarity the mutual overlaps of the
occurring cliques of the same level are not represented).
\begin{figure}[h]
\centerline{\epsfig{file=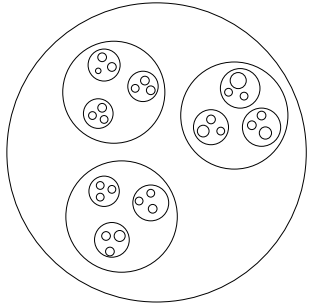,width=4cm,height=4cm,angle=0}}
\caption{Nested Structure}
\end{figure}

Each intermediate graph or array of lumps, $G_l$, carries a certain
geometric and \tit{metrical} structure of its own. We can define a
\tit{metric}, $d_l$ on $G_l$ (there exist in fact several
possibilities) as follows. We can either use the canonical graph
metric (distance of nodes measured by the minimal number of edges
connecting them) or use a more refined metric which incorporates the
varying possible degrees of overlap of cliques (cf. \cite{Req2}):
\begin{equation}d_l(S^l_i,S^l_j):=   d_{sim}(S^l_i,S^l_j):=\inf_{\gamma}\sum
  p(S^l_{k_l},S^l_{k_{l+1}})\end{equation}
where 
\begin{equation}p(A,B):=1- \operatorname{sim}(A,B)\end{equation}
and
\begin{equation}\operatorname{sim}(A,B):=[A\wedge B]/[A\vee B]\end{equation}
($\wedge,\vee$ denoting intersection and union of sets).

The above definition is understood as the infimum over the class of
paths,$\gamma$, connecting the two meta nodes in the respective
graph of $l$-cliques.\\[0.3cm]
Remark: Strictly speaking, the definition in its above form applies
only to cliques of finite order. If necessary, corresponding
definitions can be made employing measure theoretic concepts (cf.
sect. 7 of \cite{Req2}). On could of course also choose the canonical
graph distance which is, however, discrete.\vspace{0.3cm}

Concerning the importance of a true coarse-graining including an
appropriate purification, the following rigorous result is
instructive. The picture frequently invoked (\tit{space-time foam})
is the following. On a very primordial scale we have a very erratic
space-time structure having not even a stable integer dimension
(rather being of a \tit{fractal} type). Smoothing and/or
coarse-graining may ultimately lead to a a smooth continuous
space-time manifold as we know it. In \cite{dim} we introduced the
concept of (internal) dimension of a graph or network. We later
learned that it is also called the \tit{distance degree sequence} in
graph theory (cf. \cite{Small World}). In the above paper we motivated
why it should rightly be regarded as a kind of dimension and why it is
an important graph characteristic.

To put it briefly, dimension mostly enters physical models via the
asymptotic scaling of the number of degrees of freedom which can be
reached after a certain number of steps starting from a fixed
reference point. This is implemented in the following definition.
\begin{defi}[Internal Scaling Dimension] 
  Let $x$ be an arbitrary node of $G$. Let $\#(U_l(x))$ denote
  the number of nodes in $U_l(x)$, the neighborhood of nodes with
  graph distance $\leq l$ from $x$ .We consider the sequence of
  real numbers $D_l(x):= \frac{\ln(\#(U_l(x))}{\ln(l)}$. We say
  $\underline{D}_S(x):= \liminf_{l \rightarrow \infty} D_l(x)$ is the
  {\em lower} and $\overline{D}_S(x):= \limsup_{l \rightarrow \infty}
  D_l(x)$ the {\em upper internal scaling dimension} of G starting
  from $x$. If $\underline{D}_S(x)= \overline{D}_S(x)=: D_S(x)$ we say
  $G$ has internal scaling dimension $D_S(x)$ starting from $x$.
  Finally, if $D_S(x)= D_S$ $\forall x$, we simply say $G$ has {\em
    internal scaling dimension $D_S$}.
\end{defi}
Remark: For a rigorous implementation of this concept of dimension we
employ infinite graphs. For practical purposes it is of course
sufficient to have very large graphs. Note that a similar attitude is
frequently adopted in the theory blockspin renormalisation, where, if
one works with large but finite systems, the system would shrink after
every step by a certain factor. This is compensated by a rescaling of
the system.\vspace{0.3cm}

If in the preceding construction we decide not to suppress weak
(marginal) overlaps, that is, to use the unpurified ordinary clique
graph, $\hat{G}_{cl}$, we have the following remarkable result
(\cite{Req3}).
\begin{satz}Assuming that $G$ has dimension $D$ and globally bounded
  node degree we have $D_{cl}=D$. That is, the dimension does not
  change when going from a graph to its non-purified ordinary clique
  graph.
\end{satz}
This observation reminds one of a similar behavior of the
(uncoarse-grained) entropy functional in statistical mechanics which
is a constant of motion as the corresponding measure is invariant
under time evolution. Only the coarse-grained entropy happens to
increase in a non-equilibrium state. Further even more surprising
results can be found in the following section.

We now briefly indicate how one may construct a continuous space from
extracting information from our scale of networks or lump spaces. We
take the lumps of a certain renormalisation level and try to arrange
them and their mutual overlaps in some real embedding space in
essentially the same way as the corresponding cliques or lumps in the
coarse-grained graph, $G_{l_0}$, say. We discussed such geometric
constructions in much more detail in \cite{Req2}, employing, among
other things, \tit{fuzzy geometric methods}. As depicted in the
preceding picture, we endow these geometric lumps with the same nested
structure as the lumps or cliques of our coarse grained graphs, i.e.
$l_0$-cliques containing $(l_0-1)$-cliques and so forth down to the
initial nodes and bonds. In the different context of loop quantum
gravity similar constructions are discussed in \cite{Bombelli}.
Furthermore we want to mention an interesting discussion of a
hierarchy of limits of measurability in \cite{Padma}.

In making this association, it becomes obvious that there may arise
spatial \tit{obstructions} or \tit{frustrations} in case the different
links occurring in the graph $G_{l_0}$ cannot be implemented
geometrically by an appropriate \tit{packing} of overlapping balls in
a $D$ dimensional continuum. A more rigid implementation is via
\tit{simplicial complexes}. This would be a more traditional method in
which contact is mediated by having a common face. Put differently, it
is not an automatic property that the packing of these lumps fits into
some quasi smooth manifold-like structure. Certain crucial properties
like a scale free geometric long range order have to be fulfilled, see
the next section or section IV of \cite{Req3} or \cite{Small World} as
to a more detailed analysis of critical network states.

A last remark concerns the relation of the metrics or distance
functions in the two scenarios. As described above we have a couple of
natural metrics on our (clique) graph at our disposal, either discrete
ones or randomized or smooth ones. If one wants to relate such a
grainy distance function to a truely continuous version in some final
smooth space strict isometry of mappings between metric spaces is
certainly not the most natural concept. A weaker notion is
frequently (and in particular, in physics) more appropriate. Such a
concept is the notion of \tit{rough} or \tit{quasi-isometry} (see, for
example, \cite{Bridson} or \cite{Harpe}). This notion is defined as
follows.
\begin{defi}Let $F$ be a map from a metric space, $X$, to a metric
  space, $Y$ with metrics $d_X,d_Y$. It is called {\em
    quasi-isometric} if the following holds: There exist constants,
  $\lambda\geq 1,\varepsilon\geq 0$, such that
\begin{equation}\lambda^{-1}\cdot d_X(x,y)-\varepsilon\leq
  d_Y(F(x),F(y))\leq \lambda\cdot d_X(x,y)+\varepsilon \end{equation}
\end{defi}
\section{\label{translocal} The Translocal Network}
We now come to the central part of our geometric renormalisation group
analysis. Given a large not too sparsely wired network or graph, $G$,
(that is, the existing generic cliques are not too small), we
construct its canonical (unpurified) clique graph, $\hat{G}_{cl}$, and
then delete, according to our coarse graining or purification
prescription, certain bonds in $\hat{G}_{cl}$ as described above or in
\cite{Req3}.
  
Each clique or lump, $S_0$, lying in $G_{cl}$, has its own
neighborhood structure, its \tit{local group}, given by the cliques,
$S_i$, being directly connected with $S_0$ in the clique graph
$G_{cl}$, that is, having sufficient overlap with $S_0$. We can
estimate the cardinality of the typical local group of a given clique
and compare it with the total number of cliques in $G_{cl}$ or the
number of cliques, not overlapping with $S_0$. In \cite{Req3},
extensively using random graph theory, we got the following
approximate result.
\begin{conclusion}
\begin{equation}\langle N_{loc.gr.}\rangle\approx N_{cl}/(n^{l_0}\cdot\bar{r}^{\bar{r}})
 \end{equation}    
 with $n$ the number of nodes in the graph, $G$, $N_{cl}$ the number
 of generic cliques in the corresponding clique graph, $l_0$ the
 assumed sufficient degree of overlap of the generic cliques,
 $\bar{r}$ some appropriate value lying in the interval $[r_0/2,r_0]$,
 $n\gg \bar{r}\gg l_0$ being assumed (where the second $\gg$ is not so
 pronounced as the first one; $n$ is usually gigantic compared to the
 typical clique size $\bar{r}$!).
\end{conclusion}

Both $N_{cl}$ and $n$ are typically quite large in our model examples.
If $r_0$ is not too small, $l_0$ has to be chosen larger than $1$. We
conclude that in this regime most of the cliques have zero or only
marginal overlap with a given clique, $S_0$. That is, most of the
edges, occurring in the unpurified clique graph, have to be deleted
or, in other words, are only weak links (see below). Hence, already
after a single coarse-graining step, most of the fine structure
happens to be smoothed out. We therefore have the situation that, with
$p^{-1}=O(1)$, or $p$ not too small, between two arbitrary
non-overlapping cliques, $S_i,S_j$, there will nevertheless usually
exist links (of the preceding level), connecting individual nodes
lying in $S_i,S_j$ respectivly.
\begin{ob}Under the above assumptions there exist usually an
  appreciable number of nodes, $x_i,x_j$ lying in $S_i,S_j$
  respectively, such that $d_G(x_i,x_j)=1$ or, more generally,
  $d_G(x_i,x_j)$ small, while $d_{G_{cl}}(S_i,S_j)$ may be large in the
  purified clique graph $G_{cl}$.
\end{ob}

This process of coarse graining is repeated up to the level, $l_0$,
which is assumed to be sufficiently near to the macroscopic continuum.
On every step we observe this phenomenon of the existence of two types
of links between lumps or cliques. We thus get in the end a
complicated nested hierarchical structure of different types of links
between the final infinitesimal neighborhoods on the macroscopic level
and coming from all the levels below the final level, $l_0$.
Remember in particular, that, by construction, nodes on a level $l$
with $1<l<l_0$ represent full cliques on the preceding level $(l-1)$,
and by the same token, edges on level $l$ are given by non-marginal
overlaps of cliques on the preceding level.

\begin{ob}What we have described above is the two-level (or, rather,
  multi-level) structure of macroscopic space-time we alluded to in
  the introduction, that is, a continuous, locally behaving
  macroscopic space-time manifold, $M$, and an immersed web, $W$, of
  translocal (weak) links, representing, so to speak, {\em short cuts}
  between more distant regions of $M$.  Thinking of the picture of
  quotient spaces and equivalence relations, discussed in preceding
  sections, we dubbed this compound structure $QX/ST$ ($QX$ standing
  for `quantum space' and $ST$ for macroscopic space-time) in our
  earlier papers.
\end{ob}  

In the following figures we try to make this complicated and layered
structure among the nodes and cliques a little bit more transparent.
We draw, for example, two non-overlapping cliques, $S_1,S_2$, together
with some members of their local groups. The euclidean distance in the
picture is meant to indicate their large relative distance in
$G_{cl}$. We assume that two nodes, $x,y$, lying in $S_1,S_2$
respectively, are connected by a link belonging to $G$.

In any general clique graph we have this difference between weak and
strong links. If our graph is a fortiori a sufficiently typical random
graph this structure is even a little bit more particular as we
learned above that in a typical random graph practically all cliques
are almost of the same size (their order lying in the interval
$r_0/2,r_0$). That is, each weak link belongs actually to at least one
other clique of practically the same order.
\begin{bem}This last property may be different in graphs, having a more
  pronounced \tit{near-} and \tit{far-order} as discussed in
  \cite{Small World} and further literature given there.
\end{bem}
In our illustration 
\begin{figure}[h]
\centerline{\epsfig{file=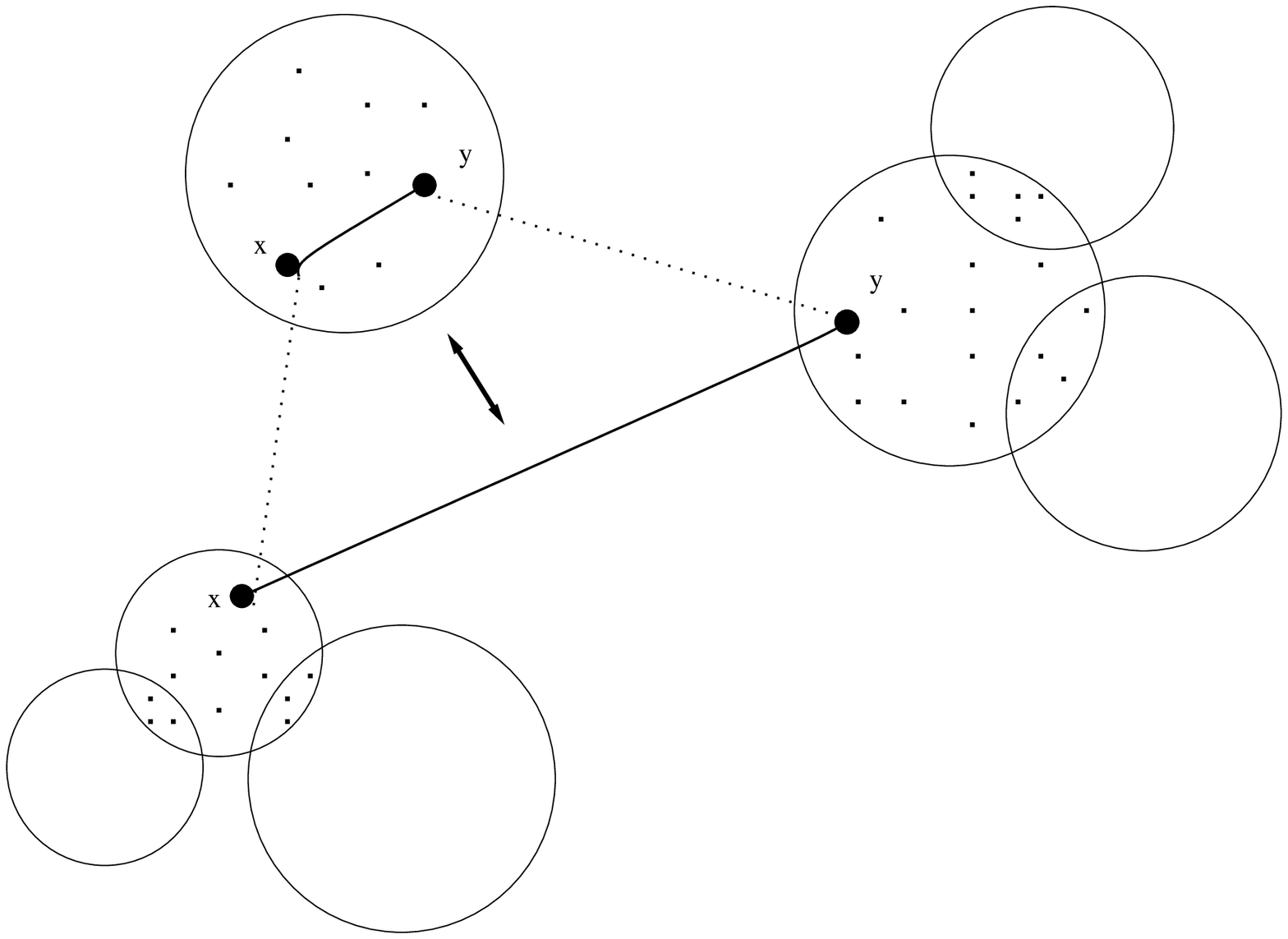,width=6cm,height=6cm,angle=0}}
\caption{Translocal Links 1}
\end{figure}
the pointed lines plus the arrow mean that the points $x,y$, occurring
twice, have to be identified and the corresponding lines to be
contracted. That is, the third clique has actually a common node both
with $S_1$ and $S_2$. A connection  via an intermediate clique is
one possibility. The following picture describes a direct (weak)
contact of $S_1$ and $S_2$ via a single common node $x$.
\begin{figure}[h]
\centerline{\epsfig{file=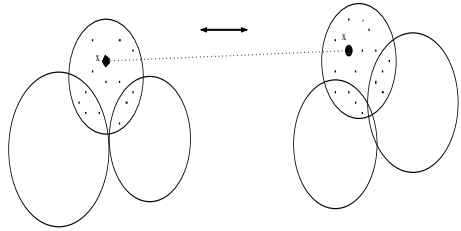,width=6cm,height=3cm,angle=0}}
\caption{Translocal Links 2}
\end{figure}

Up to now we arrived at our conclusions concerning the translocal
structure of our coarse-grained macroscopic limit networks mainly with
the help of the random graph framework plus certain consequences of
the geometric renormalisation process. We will now amend these
observations with some precise results which almost rigorously show
that these limit networks have to be of a \tit{critical scale-free
  small world} type.

One of our main conceptual tools will be the behavior of the dimension
of a network under coarse-graining. We showed in the preceding section
that coarse-graining is absolutely necessary if we want to change or
reduce the presumably (fractal) dimension of the initial network on
the primordial Planck scale. As a corollary of the above theorem we
have:
\begin{koro}For the purified clique graph (in contrast to the
  unpurified one), we have
\begin{equation}D_{cl}\leq D\quad\text{instead of}\quad D_{cl}=D \end{equation}
\end{koro}

One may be inclined to surmise that generically we will have
$D_{cl}<D$, but this is not true! Quite to the contrary, it turns out
to be very tricky to really reduce the dimension. \tit{Local}
alterations of the wiring diagram will not do. Analysing the situation
in which a dimensional reduction can actually take place leads to the
concept of \tit{critical network states}. We approach the problem of
dimensional reduction by smoothing in two steps. We learned that
\begin{equation}D_G=D_{\hat{G}_{cl}}\leq D_{G_{cl}}\end{equation}
That is, it is sufficient to controll the step from $\hat{G}_{cl}$ to
$G_{cl}$. This transition consists in the deletion of a a certain
fraction of (weak) links.

In \cite{dim}, sect. 4.1 we proved an interesting theorem which shows
that the local insertion of arbitrarily many additional edges does not
change the dimension of a graph. More precisely: 
\begin{propo}Additional insertions of bonds between arbitrarily many nodes,
  $y,z$, having original graph distance, $d(y,z)\leq k\;,\;k\in\N$
  arbitrary but fixed, do not change $\underline{D}(x)$ or
  $\overline{D}(x)$.
\end{propo}
Whereas edge deletion is not simply the dual operation of edge
insertion, we can generalize the above result in a way appropriate for
our problem. What we need is a generalisation of the notion
\tit{local}. 

Both $\hat{G}_{cl}$ and $G_{cl}$ carry their own natural graph
metrics, $\hat{d}$ and $d$, given by the \tit{geodesic} edge path
distance. 
\begin{defi}We assume that we pass over from a graph G to a new graph
 $G'$, living on the same node set, by means of a number of edge
  deletions. These edge deletions are called local of order $\leq k$
  if only edges between nodes, $x,y$, are deleted which have a final
  distance in $G'$ globally bounded by $k$.
\end{defi}
We then have the remarkable theorem:
\begin{satz}If the edge deletions in going from $\hat{G}_{cl}$ to
  $G_{cl}$ are uniformly local of an arbitrary but finite order, the
  dimension does not change, i.e.
\begin{equation}D_{\hat{G}_{cl}}=D_{G_{cl}}    \end{equation}
\end{satz}
Proof: This can be proven by reversing the proof of the above cited
proposition, i.e. we envisage the dual process of going from $G_{cl}$
to $\hat{G}_{cl}$ by edge insertions, which are now, by assumption,
local of a certain finite order. From our proposition we now can infer
that the dimension remains unchanged.\bewende
\begin{koro}In order to change the dimension by edge deletions (in an
  infinite graph; see the remarks after definition 5.3) it is
  therefore necessary (but not sufficient; there are counter
  examples!) that there exist infinitely many edge deletions with
  their degree of locality not being boundable by any given number k.
  Put differently, for any given k there exist infinitely many edge
  deletions between nodes with final distance greater than k.
\end{koro}
\begin{conclusion}We view this kind of scale-free behavior of the
  occurrence of long-distance weak links as a rigorous formulation of
  the critical scale-free network states mentioned in the
  introduction.
\end{conclusion}  

We want to complement these general results with an illustrative
example, which shows all the features we mentioned above. We emphasize
however that it is \tit{not} meant as an example of a limit state of a
true proto space-time of the kind we have discussed in the preceding
sections. It is only a toy model! In a first step we embed the
one-dimensional line of integers, $\Z_1$, in the two-dimensional
lattice, $\Z_2$, in the way depicted in the following figure.
\begin{figure}[h]
\centerline{\epsfig{file=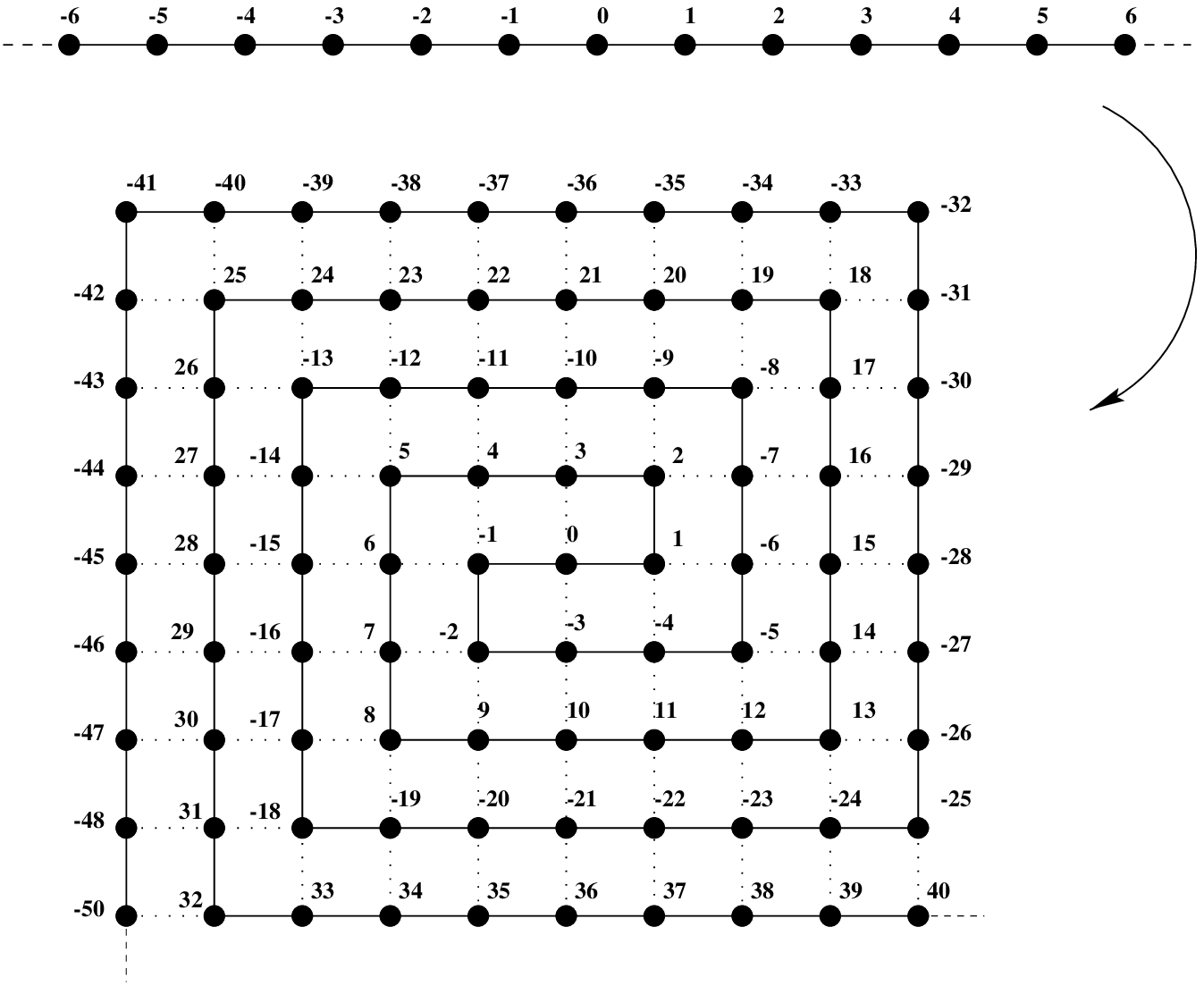,width=6cm,height=6cm,angle=0}}
\caption{}
\end{figure}
We regard this ambient lattice as the unpurified clique graph,
$\hat{G}_{cl}$, of a certain coarse-graining level, with the nodes
being cliques. Strong links, given by a sufficient overlap, are
denoted by bold lines, weak bonds, which have to be deleted when
going from $\hat{G}_{cl}$ to the purified clique graph, $G_{cl}$, by
dashed lines.

We now see that $\hat{G}_{cl}$ has graph dimension two while $G_{cl}$
has dimension one. We learned from our previous rigorous results that
such a change can only occur if $\hat{G}_{cl}$ is in a very peculiar
critical state. Inspecting our toy model we see that exactly this is
the case (what is actually peculiar is the embedding of $\Z_1$ in
$Z_2$). Node distance in $G_{cl}$ is measured by the canonical metric
of $\Z_1$, while the metric in $\hat{G}_{cl}$ is the one coming from
$\Z_2$. We see that the bond deletion process really violates the
locality assumption for any given $k$. For any given $k$ there exist
an infinity of weak links (to be deleted), connecting nodes having a
distance larger than $k$. Take for example the following sequence of
weak links (ordered by increasing node distance)
\begin{equation}(0,3)\;,\;(3,-10)\;,\;(-10,21)\;,\;(21,-36)\ldots\end{equation}
where the numbers denote the position of the nodes with respect to $\Z_1$. 
\section{Wormhole Spaces or a Continuum Model of Points Speaking to Each Other}
In the first sections of the present paper we discussed a general
point of view concerning a wider conception of continuous spaces,
being of possible relevance for (quantum) space-time physics.  We then
discussed the subject from a different angle, i.e., as dynamical,
densely entangled networks of relations among microscopic
constituents, being depicted by nodes, the relations or elementary
interactions by edges. If we perform a sequence of specific
coarse-graining steps on this network, which, under certain
conditions, will finally converge to a smooth macroscopic space or
space-time, a detailed analysis shows the following. With the help of
the random graph concept we observe the quasi automatic emergence of a
new and subliminal web of \tit{translocal} interactions, being
immersed in this classical manifold, $M$. In the following subsections
we want to give a brief account of the sort of continuous model spaces
we expect to emerge from this construction, that is, ordinary
continous spaces with an embedded web of translocal \tit{short cuts}.
We introduce these spaces in a quasi axiomatic way as it may sometimes
be easier to start right away from these models, if one wants to work
out the consequences for continuum physics (like, for example ordinary
quantum theory) instead of going the long way beginning at the Planck
scale. Confer also the illuminating remarks by DeWitt cited in the
introduction.
\subsection{A Class of Continuum Models}
We conclude that these spaces which, presumably, emerge in quantum
space-time physics, support two modes (or rather a whole hierarchy) of
interactions and/or information exchange among their constituents. A
\tit{local} one, obeying the ``\tit{Nahwirkungsprinzip}'' (\tit{no
  action at a distance}), propagating from (physical) points (or
lumps) to their \tit{infinitesimal neighbors} and so on, and, on the
other hand, a \tit{translocal} almost quasi-instantaneous (but
presumably feeble) interaction with arbitrarily distant regions of the
manifold $M$ of a more stochastic type.

This presumed more irregular and \tit{stochastic} behavior is a result
of the weak contact (via the weak links) between individual nodes in
the various lumps making up the physical points being translocally
related in contrast to the more robust interaction (strong local
links) given by a more intense overlap of full cliques or lumps as is
the case in the infinitesimal neighborhood or (in the clique or lump
language) in the local group. One can study this different behavior
when imposing a dynamical microscopic law as we discussed it in
previous papers. One usually observes large individual fluctuations at
individual nodes or along individual links of the primordial level
which are then smoothed and averaged out over full cliques or clusters
of cliques on the higher levels of the network. This is now the place
where our line of argumentation returns to the point we departed from,
that is, the picture of ``\tit{points talking to each other}'', which
we invoked in the first sections.

To begin with, the implementation of our findings by means of a
continuum description is surely not unique! From a mathematical point
of view a whole \tit{class} of spaces, all sharing certain basic
characteristics, can be invented. So we begin by introducing some
models which do not yet share \tit{all} the features we expect from
our renormalisation group analysis.

We start with some continuous space, $M$, like e.g. $\R^d$, or a
manifold, being locally homeomorphic to some $\R^d$. We assume that in
$M$ a countable but dense subset, $X$, or, alternatively, a partition
of non-overlapping, countable and dense subsets, $X_{\iota}$, is
specified:
\begin{itemize}
\item $X$ or $X_{\iota}$ are countable and dense in $M$ (note that they
  do not contain interior points with respect to the topology of $M$).
\item $X_{\iota}$ do not overlap and $\cup X_{\iota}=M$. In another
  model situation we may assume that the $X_{\iota}$ are not dense
  with $\cup X_{\iota}\neq M$ but dense in $M$.
\end{itemize}
The above assumptions describe slightly different models and there
certainly do exist more model systems of this kind. We surmise however
that, on a more macroscopic scale, the correct choice is perhaps not
really crucial. Central is the idea that these delocalized sets should
somehow be meager compared to the full continuum, that is, they should
have Lebesgue measure zero. On the other hand, sets of the fractal
type (cantor dust) may be admissible.\vspace{0.3cm}

As we invoked in previous sections the renormalisation group picture,
the phenomenon called \tit{universality} comes to mind. There may in
fact exist different microscopic model systems all converging to the same
coarse grained macroscopic fixed point provided that they share
certain crucial characteristics, determining the whole class. In our
case this is the particular kind of non-local \tit{entanglement}. 

In the `foliation-model' (alluding to a situation similar to the
non-commutative torus, discussed above), i.e. all $X_{\iota}$ dense in
$M$ but non-overlapping, we encounter the following situation.  Every
point of $M$ belongs to exactly one of the subsets, $X_\iota$.  Each
of these subsets is spread over the whole manifold $M$ and we have in
particular that for each neighborhood, $O_y$, of some point, $y\in M$
\begin{equation}X_\iota\cap O_y\;\text{dense in}\;O_y\end{equation}
\begin{ob}The above partition defines an equivalence relation
  $R\subset M\times M$ with members $(x,y)$ so that $y\in \tilde{x}$
  with $\tilde{x}$ the set $X_\iota$, $x$ is belonging to. As in the
  introductory sections, we can define a bundle structure,
  $\tilde{M}$, with base space $M$ and fibers $(x,\tilde{x})$ and
  proceed in the same way as above by e.g. introducing the local
  Hilbert spaces, $H(\tilde{x})$ which are attached as fibers, $H_y$
  to the members, $y$, of the equivalence class. Correspondingly, we
  can introduce random operators acting in these local Hilbert space
  fibers.
\end{ob}
\begin{conclusion} As in Connes approach we see, that the natural
  structure is not really some kind of quotient space but rather an
  amplification of the original space, $M$, to a fiber space over $M$.
  The internal spaces describe the subset of points which ``can speak
  to each other'' in a translocal way. These subsets are the classes
  of points of $M$, which happen to be connected by weak links in the
  underlying network, as has been descibed in the preceding section.
\end{conclusion} 

We already remarked in section 2 that the final continuum structure,
expected to emerge from our primordial dynamical network, may be more
special what concerns the embedded translocal web than a simple
equivalence relation (see also \cite{Req3} and, in particular the
section about networks as \tit{causal sets}). As in the dynamical
laws, employed by us, the edges typically carry states labelled by
$(\pm 1,0)$ being associated with the two possible orientations of the
edge or the non-active state, we have rather a relational structure
corresponding to a directed or oriented graph.

Furthermore, the ordinary mathematical model spaces are \tit{static}.
As we discussed in earlier papers, the states on the network and hence
also the geometric structure follow a dynamical network law which
constantly changes the wiring, the orientation of the edges and the
shape of the cliques or lumps together with their mutual overlap. We
took this into account in \cite{Req2} by emulating it on the more
macroscopic levels in form of \tit{fuzzyness} of shapes and
\tit{randomness} of, for example, distances. That is, for each point
$x\in M$ we rather have a (dense) set of distant points, $[x]_{in}$,
sending information to $x$ and another set, $[x]_{out}$, getting
information from the point $x$, both sets being time dependent.

Note that in contrast to equivalence relations, subsets like
$[x]_{in}$ or $[x]_{out}$ do not lead to a partitioning of $M$. There
can exist complicated overlap patterns for the respective sets
belonging to points $x\neq y$, nor does there exist a transitivity
relation. On the other hand, on a macroscopic level, the differences
between the various model spaces of the \tit{wormhole class} to be
found on the finer scales may be washed out. As all the above subsets
are expected to be dense in $M$ or $X$, and therefore also in each
given region of $M$, any region $O_1$ gets translocal (stochastic)
information from any other region $O_2$ and vice versa.
\subsection{Microscopic Wormholes and Wheeler's Space-Time Foam}
The chain of thoughts, presented in the preceding sections, leads to a
new microscopic picture of space-time and/or the \tit{quantum vacuum},
strongly suggesting a \tit{translocal entanglement} among distant
regions of our continuous manifold. This structure is encoded in a web
of relations which is largely hidden on the surface level of
(quasi)classical space-time but which, as we think, becomes observable
through its expression in various features of quantum non-locality
(cf. the remarks of v.Weizsaecker cited in the introduction).

So far our approach was decidedly bottom-up, starting from a presumed
underlying microscopic substratum and reconstructing the more
macroscopic levels by a renormalisation-like process of
coarse-graining. On the other hand, there does exist for already quite
some time a more top-down oriented picture, which, coming down from
the continuum side of physics, envokes the scenario of a foam-like
substructure of space-time on the Planck scale. In this context
Wheeler et al developed the idea of microscopic wormholes, connecting
distant parts of our ordinary space-time manifold or even different
universes (see e.g.  the classical book by Misner, Thorne, Wheeler;
\cite{Wheeler1}). A beautiful and more up to date presentation can be
found in \cite{Fisser}.

We note that both ideas are realized in our framework. The idea of a
foamy space which is almost fractal on a truely microscopic scale and
has a scale or resolution dependent dimension is realized in our
geometric renormalisation procedure provided that the network is in a
critical state having a dense web of practically scale-free
translocal links. This was rigorously shown in the preceding
sections. It is suggestive to associate these translocal links with
the presumed microscopic wormholes of Wheeler.
\begin{ob}Associating our web of translocal links with the microscopic
  wormholes of Wheeler we have shown that there existence is crucial
  in order that the dimension of space-time can become a scale
  dependent property, decreasing from a presumably large (fractal)
  microscopic dimension to the small integer dimension of ordinary
  classical space-time. That is, both ideas belong closely together.
\end{ob} 

We remark that the scenario we are envisaging is not so far-fetched as
it may seem. There exist, in fact, several recent investigations
concerning the possible role of wormholes for the stability of the
ordinary vacuum in quantum gravity. The possible effects of a gas of
Planckian wormholes on various physical phenomena were studied several
times in the past; as an example we mention the paper by Coleman
(\cite{Coleman}). In \cite{Preparata} it was argued, that in quantum
gravity an array of Planckian wormholes may be the correct ground
state. This short list is far from being complete.  All these
speculations and observations seem to underpin our own line of
reasoning.
\section{A Brief Outlook on Quantum Entanglement and Other Translocal
  Quantum Phenomena} One of our motivations, to develop the above
framework, is the goal to reach a better (and more realistic)
understanding of the many mysteries being inherent in the various
phenomena of \tit{quantum non-locality} and \tit{entanglement}, the
evident, but not well understood, necessity of \tit{complex}
superposition, \tit{interference} and the peculiarities of the
\tit{measuring process}. Some of these points have been already
discussed in a preliminary form in \cite{Quantum}. Now, with the
concept of wormholespaces at our disposal, we are able to analyze
these phenomena with greater rigor. In order not to blow up the size
of the present paper beyond reasonable length, we will however make
only some general remarks.

There exist several papers in the more recent past, which strike a
similar key as far as the general working philosophy or parts of the
present analysis are concerned (while the technical framework may be
quite different). The following brief remarks are not meant as a full
discussion of the field. We mention only a few points of view which
seem to be particularly close to our own approach. An interesting
approach has been developed by Smolin
(\cite{Smolin1},\cite{Smolin2},\cite{Smolin3}). It is perhaps
intriguing to relate the \tit{matrix-model} approach in the latter
paper to our \tit{bundle} or \tit{foliation structure}. In both cases
we have an array of countable subspaces which interact with each
other. At the end of \cite{Smolin2}, on the other hand, one can find a
brief discussion of a relational
description of space-time in form of graphs.

A technically slightly different line of ideas is pursued in the
following papers of 't Hooft
(\cite{Hooft1},\cite{Hooft2},\cite{Hooft3}). In this approach a
deterministic \tit{cellular automaton}-like primordial substratum is
introduced which is similar to but more regular and static than our
dynamical cellular network, $QX$. It is argued that quantum theory
might emerge on a larger scale from such a derministic and regular
array. This approach has also  been briefly discussed by us in
\cite{Quantum}.

Before we come to the more obvious consequences of our presumed
wormhole structure we want to briefly mention that we think that the
(somewhat mysterious) need of employing a \tit{complex} superposition
principle and a \tit{complex structure} in general (\cite{Asht}) is
also a (subtle and not so obvious) consequence of this additional
translocal web, being embedded in the ordinary continuous and locally
behaving space structure. 

It is quite funny in this respect to see that Schroedinger himself, in
his row of five or six epochal papers about wave mechanics, for a very
long time hold the view that, as in ordinary classical undulatory
physics, one can always go over to the real part of the complex wave
function if one wants to. Only in the last paper of this series
(\cite{Schroedinger1}) it began to dawn on him (possibly inspired by a
letter from Lorentz) that the complex structure of quantum theory is
inevitable and is buried deeply in its foundational structure (see
also the beautiful essay by Yang \cite{Yang}). The possible
consequences and deficits of quantum mechanics over a \tit{real}
Hilbert space were analyzed in some detail in \cite{Stueckel}, see
also the remarks in section 8 of \cite{Jauch}.  Very illuminating in
this respect is the paper of Dirac (\cite{Dirac}) in which he rightly
argues that the emergence of a \tit{phase quantity} is presumably even
more important than the emergence of non-commuting observables.

More obvious is the effect which the translocal web of weak bonds will
have on the understanding of the quantum mechanical measurement
process, on \tit{entanglement} and other related phenomena. It was
exactly the phenomenon of seemingly \tit{instantaneous collapse} which
stood in the way of a more \tit{realistic} interpretation of the
extended complex wave pattern.

Assuming that Einstein causality also holds sway in the quantum
regime, quasi-instantaneous destruction of those parts of the wave,
being located outside the region of direct measurement interference,
could only be explained by granting the wave function, or more
generally, the quantum state only the ontological status of a mere
\tit{bookkeeping device} of the \tit{(non)-knowledge} of the observer.
Looked upon from a slightly different angle, this explains the
dominance of the \tit{ensemble picture}.

On the other hand, if, in addition to the ordinary \tit{local} and
\tit{causal} propagation between neighboring lumps and taking place
with a \tit{finite} velocity, we have a further, more subliminal
\tit{translocal} information transport through the web of weak bonds
or, in more popular terms, through \tit{hyperspace}, the almost
instantaneous destruction of a real and existing excitation pattern of
the vacuum becomes possible. Similar considerations hold in the
context of entanglement. The details of these processes need of
course a subtle analysis.

As a last point to mention we want to briefly comment on the seeming
dichotomy between the presumed underlying and almost hidden translocal
substructure and, on the other hand, the apparent local representation
in form of partial differential equation. It is one of our findings
that such non-local contributions and effects may come in a
\tit{local} disguise, such that, without an underlying more
fundamental theory, it turns out to be difficult or nearly impossible
to detect the translocal pieces of a model theory and separate them
from the local ones. To give an example what we have in mind. An
observer, prepared to take only local interactions into account, may
formulate an effective theory of locally interacting fields,
$A_i(x,t)$. However, these seemingly local fields may, for example,
represent $(x,t)$-dependent integrals over distant contribution of the
field configuration or even over quantities, which do occur on a
deeper, more primordial level (as discussed in our coarse graining
process). It could in particular happen that coupling constants turn
out to be such integrated non-local quantities, an idea which
obviously carry a strongly Machian spirit. To show that quantum theory
would exactly be such an effective theory will represent the next
logical step.

\end{document}